\documentclass[%
  reprint,
  showpacs,preprintnumbers,
  amsmath,amssymb,
  aps,
  prb,
]{revtex4-2}

\usepackage{graphicx}[]
\usepackage{dcolumn}
\usepackage{bm}
\usepackage{multirow}
\usepackage{braket}
\usepackage{color}
\usepackage{ulem}

\usepackage{hyperref}
\usepackage{url}

\begin{document}

\preprint{APS/123-QED}

\title{
Skyrmion crystal with integer and fractional skyrmion numbers\\ in a nonsymmorphic lattice structure with the screw axis
}

\author{Satoru Hayami}
\affiliation{
Department of Applied Physics, The University of Tokyo, Tokyo 113-8656, Japan
}
 
\begin{abstract}
The emergence of a magnetic skyrmion crystal in a nonsymmorphic lattice system with the screw symmetry is numerically investigated. 
By performing the simulated annealing for a layered spin model with the isotropic exchange interaction and the antisymmetric Dzyaloshinskii-Moriya interaction, and then constructing a low-temperature phase diagram, we reveal that the skyrmion crystal is stabilized in both zero and nonzero fields even without the threefold rotational symmetry in the two-dimensional plane. 
We show that a competition between the ferromagnetic interlayer exchange interaction and the layer- and momentum-dependent anisotropic interactions is a source of the skyrmion crystals in the presence of the screw axis. 
Moreover, we find two types of layer-dependent skyrmion crystals, which are characterized as a coexisting state of the skyrmion crystal and the spiral state, in the narrow field region.  
Our result provides a reference for further search of the skyrmion crystals in nonsymmorphic lattice systems. 

\end{abstract}
\maketitle

\section{Introduction}

Noncollinear and noncoplanar spin configurations have drawn considerable attention in condensed matter physics, as they give rise to intriguing collective excitations, multiferroic phenomena, and transport properties. 
The microscopic key ingredients in these phenomena are often described by the vector and scalar spin chiralities, which correspond to the two-spin vector and three-spin scalar products, respectively~\cite{kawamura1984phase,lacroix2011introduction,batista2016frustration}. 
For example, the vector chirality in noncollinear magnets leads to polar-vector-related physical phenomena without spatial inversion symmetry, such as the electric polarization~\cite{Mostovoy_PhysRevLett.96.067601,cheong2007multiferroics,Bulaevskii_PhysRevB.78.024402}, the spin current generation~\cite{Katsura_PhysRevLett.95.057205,Zelezny_PhysRevLett.119.187204,zhang2018spin}, the magnetoelectric effect~\cite{seki2012observation,white2012electric,okamura2013microwave,Mochizuki_PhysRevB.87.134403,tokura2014multiferroics,Hayami_PhysRevB.90.024432,mochizuki2015dynamical,thole2018magnetoelectric,Gobel_PhysRevB.99.060406,Hayami_PhysRevB.104.045117,Hayami2022_1}, and the antisymmetric spin-split band structure~\cite{Hayami_PhysRevB.101.220403,Hayami_PhysRevB.102.144441,Hayami_PhysRevB.105.024413}. 
On the other hand, the scalar chirality in noncoplanar magnets leads to axial-vector-related physical phenomena without time-reversal symmetry, such as the topological Hall effect~\cite{Ye_PhysRevLett.83.3737,Ohgushi_PhysRevB.62.R6065,tatara2002chirality,Nagaosa_RevModPhys.82.1539,Martin_PhysRevLett.101.156402,Lee_PhysRevLett.102.186601,Neubauer_PhysRevLett.102.186602,nagaosa2013topological,hayami_PhysRevB.91.075104,Hamamoto_PhysRevB.92.115417}. 
Moreover, it was recently shown that noncoplanar magnets induce nonreciprocal transport phenomena arising from the asymmetric band deformation in terms of $+\bm{k}$ and $-\bm{k}$ ($\bm{k}$ is the wave vector) even without the spin-orbit coupling~\cite{Hayami_PhysRevB.101.220403, Hayami_PhysRevB.102.144441,hayami2021phase, Hayami2022_2}. 
In this way, materials with noncollinear and noncoplanar spin configurations provide a fertile playground, which might be potentially applied to next-generation spintronic devices. 

Among them, a magnetic skyrmion, which is characterized by a periodic array of two-dimensional topological objects, has been extensively studied in both theory and experiment~\cite{nagaosa2013topological, fert2017magnetic, Tokura_doi:10.1021/acs.chemrev.0c00297,hayami2021topological}. 
A magnetic skyrmion crystal (SkX), which is defined as a periodic array of the magnetic skyrmion, has been ubiquitously identified in various noncentrosymmetric lattice structures since its discovery in chiral cubic magnets like MnSi~\cite{Muhlbauer_2009skyrmion} and Fe$_{0.5}$Co$_{0.5}$Si~\cite{yu2010real}. 
Recently, it was also studied in different magnetic systems, such as the centrosymmetric system~\cite{kurumaji2019skyrmion,hirschberger2019skyrmion,khanh2020nanometric,Yasui2020imaging,Hirschberger_10.1088/1367-2630/abdef9,khanh2022zoology} and van der Waals system~\cite{tong2018skyrmions,ding2019observation,wu2020neel,Akram_PhysRevB.103.L140406,Ray_PhysRevB.104.014410}. 
Accordingly, a variety of underlying mechanisms to stabilize the SkXs have been found in both the localized spin and itinerant electron models. 
The uniform Dzyaloshinskii-Moriya (DM) interaction~\cite{dzyaloshinsky1958thermodynamic,moriya1960anisotropic,rossler2006spontaneous,Yi_PhysRevB.80.054416}, the staggered DM interaction~\cite{Hayami_PhysRevB.105.014408,lin2021skyrmion}, frustrated exchange interaction~\cite{Okubo_PhysRevLett.108.017206,leonov2015multiply,Lin_PhysRevB.93.064430,Hayami_PhysRevB.93.184413,batista2016frustration,Lin_PhysRevLett.120.077202,Hayami_PhysRevB.103.224418,Hayami_unpub}, and dipolar interaction~\cite{Utesov_PhysRevB.103.064414,utesov2021mean}, are typical examples for the former, while the Ruderman-Kittel-Kasuya-Yosida interaction~\cite{Ruderman,Kasuya,Yosida1957,Wang_PhysRevLett.124.207201,Mitsumoto_PhysRevB.104.184432,mitsumoto2021skyrmion}, multiple spin interaction~\cite{Ozawa_PhysRevLett.118.147205,Hayami_PhysRevB.95.224424,Hayami_PhysRevB.99.094420,hayami2020multiple,Eto_PhysRevB.104.104425,Hayami_10.1088/1367-2630/ac3683,wang2021skyrmion}, and anisotropic exchange interaction~\cite{Hayami_PhysRevLett.121.137202,amoroso2020spontaneous,Hayami_PhysRevB.103.054422,yambe2021skyrmion,amoroso2021tuning,kato2022magnetic} are ones for the latter. 

The formation of the SkXs is described by a superposition of multiple spin density waves (multiple-$Q$ state). 
In particular, a triangular SkX consisting of a triple-$Q$ spiral superposition, whose ordering vectors lie on a two-dimensional plane satisfying $\bm{Q}_1+\bm{Q}_2+\bm{Q}_3=\bm{0}$, often appears even in cubic-lattice~\cite{Binz_PhysRevLett.96.207202,Binz_PhysRevB.74.214408,Butenko_PhysRevB.82.052403,hayami2021field} and square-lattice~\cite{Yi_PhysRevB.80.054416,Lin_PhysRevB.93.064430} systems. 
This is because the above condition regarding multiple-$Q$ ordering vectors gives rise to an effective coupling as $(\bm{S}_{\bm{0}}\cdot \bm{S}_{\bm{Q}_1})(\bm{S}_{\bm{Q}_2}\cdot \bm{S}_{\bm{Q}_3})$ in the free energy, where $\bm{S}_{\bm{Q}_\nu}$ ($\nu=0$-$3$ and $\bm{Q}_{\bm{0}}=\bm{0}$) is the Fourier transform of localized spin at site $i$, $\bm{S}_i$. 
Indeed, a square SkX, which is characterized as a double-$Q$ spiral state formed by two orthogonal ordering vectors, i.e., $\bm{Q}_1+\bm{Q}_2 \neq \bm{0}$, has been stabilized by taking into account a relatively large fourfold anisotropy~\cite{Christensen_PhysRevX.8.041022}, such as bond-dependent anisotropy~\cite{Hayami_doi:10.7566/JPSJ.89.103702,Hayami_PhysRevB.103.024439,Utesov_PhysRevB.103.064414,Wang_PhysRevB.103.104408} and interaction at higher-harmonic ordering vectors like $\bm{Q}_1+\bm{Q}_2$~\cite{hayami2022multiple}. 
Reflecting such a difference, the materials hosting the square SkX like Co$_{10-x/2}$Zn$_{10-x/2}$Mn$_x$~\cite{tokunaga2015new,karube2016robust,karube2018disordered,Karube_PhysRevB.102.064408,henderson2021skyrmion}, Cu$_2$OSeO$_3$~\cite{chacon2018observation,takagi2020particle}, GdRu$_2$Si$_2$~\cite{khanh2020nanometric,Yasui2020imaging,khanh2022zoology}, and EuAl$_4$~\cite{Shang_PhysRevB.103.L020405,kaneko2021charge,Zhu_PhysRevB.105.014423,takagi2022square} are rather rare compared to ones hosting the triangular SkX. 
Thus, the lattice systems with the threefold rotational axis, such as the trigonal, hexagonal, and cubic lattice systems, provide an appropriate situation to stabilize the SkX, since the threefold rotational symmetry naturally leads to $\bm{Q}_1+\bm{Q}_2+\bm{Q}_3=\bm{0}$. 

Meanwhile, the above argument has implied that the three ordering wave vectors, $\bm{Q}_1$, $\bm{Q}_2$, and $\bm{Q}_3$, are connected by the threefold rotational symmetry within the same lattice plane. 
Then, one might wonder what happens these ordering wave vectors are related to the threefold screw axis in a nonsymmorphic lattice system, which is represented by the space groups $P3_1$ (\# 144), $P3_2$ (\# 145), $P3_1 12$ (\# 151), $P3_1 21$ (\# 152), $P3_2 12$ (\# 153), and , $P3_2 21$ (\# 154); the system does not have threefold rotational symmetry in a two-dimensional space but it is invariant under a combined operation by threefold rotation and translation. 
Although such a situation satisfies the condition of $\bm{Q}_1+\bm{Q}_2+\bm{Q}_3=\bm{0}$ in a unit cell, it is unclear whether the SkX is stabilized as a lowest-energy state.  

In the present study, we investigate instability toward the SkXs in a nonsymmorphic lattice system with a threefold screw axis. 
For that purpose, we consider a layered triangular lattice, where each layer has anisotropic interactions in different bond directions. 
After constructing a minimal effective spin model with the momentum-resolved anisotropic interactions including the DM interaction, we examine a low-temperature phase diagram by performing the simulated annealing. 
As a result, we show that the SkX is stabilized from zero to finite magnetic fields in the systems with the threefold screw axis but without the threefold rotation axis. 
We discuss the stability of the SkX based on the interplay between the anisotropic exchange interaction in momentum space and interlayer exchange interaction.  
Furthermore, we find two new types of SkXs characteristics of a threefold-screw system, where the SkX spin texture emerges in the layer-dependent form resulting in the fractional skyrmion number in a magnetic unit cell. 
Our results indicate an emergent SkX in the nonsymmorphic lattice systems with the screw axis, which gives a guideline to search for further SkX-hosting materials in a variety of lattice structures. 

This paper is organized as follows. 
First, we present a setup to investigate the SkX in the nonsymmorphic lattice system in Sec.~\ref{sec: Setup}. 
We introduce an effective spin model with the exchange and DM interactions in momentum space. 
We also outline the numerical method by using simulated annealing. 
Next, we show the magnetic phase diagram and discuss the instability toward the SkX in Sec.~\ref{sec: Instability toward skyrmion crystal phase}. 
In Sec.~\ref{sec: Skyrmion crystal phases with fractional skyrmion numbers}, we reveal that the SkXs with fractional skyrmion numbers appear in the vicinity of the SkX with the skyrmion number of one. 
Finally, the summary of this paper is given in Sec~\ref{sec: Summary}.

\section{Setup}
\label{sec: Setup}

\begin{figure}[t!]
\begin{center}
\includegraphics[width=1.0 \hsize ]{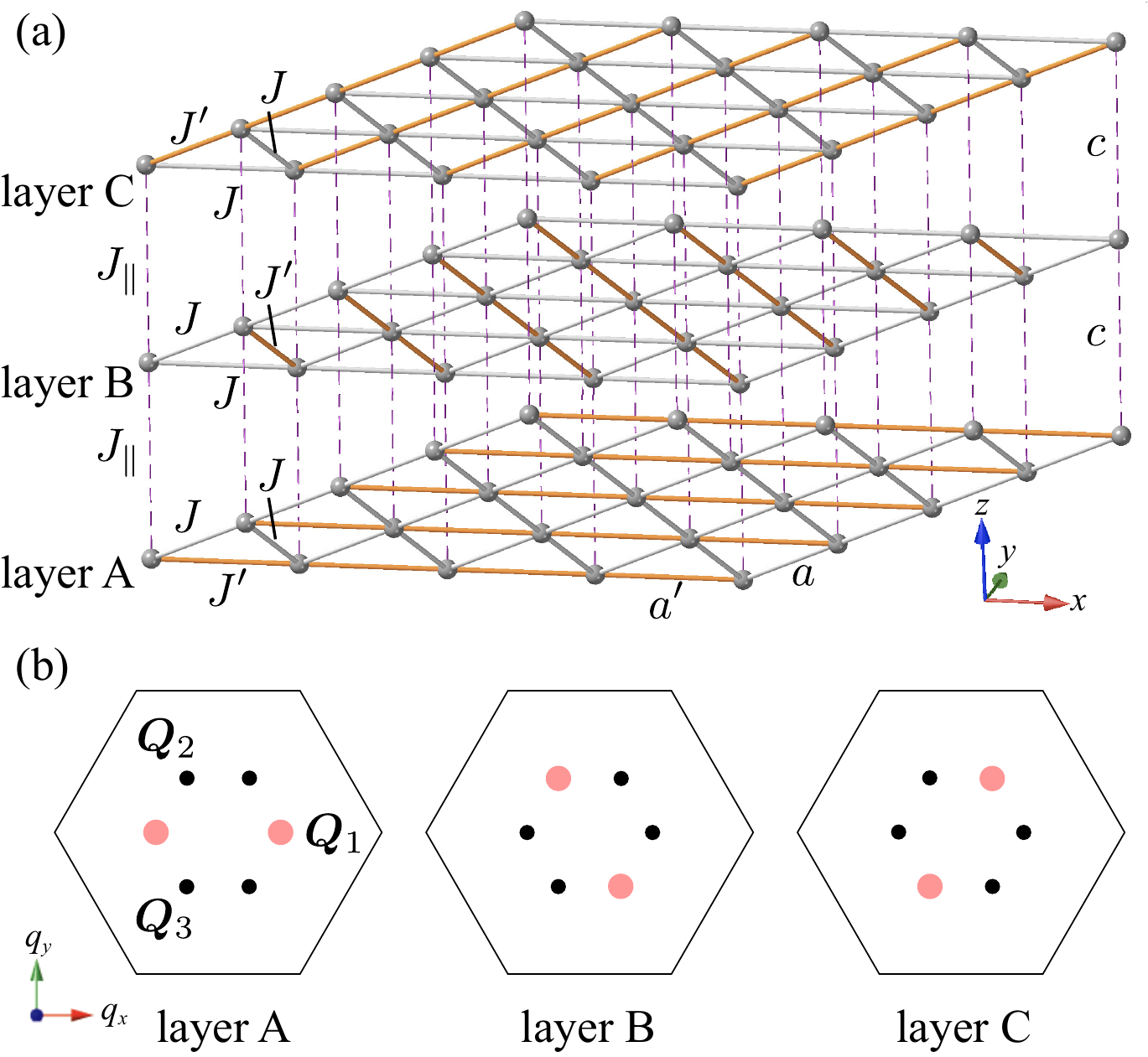} 
\caption{
\label{fig: setup}
(a) Three layers (A, B, and C) connected by the threefold screw symmetry. 
(b) Schematics of the six relevant ordering wave vectors, $\pm \bm{Q}_1$, $\pm \bm{Q}_2$, and $\pm \bm{Q}_3$, in the model in Eq.~(\ref{eq: Ham_perp2}). 
The large (small) circles represent the dominant (subdominant) components in the momentum-resolved interactions. 
}
\end{center}
\end{figure}

To investigate the instability toward the SkX in a nonsymmorphic system with a threefold screw axis,  we consider the layered triangular lattices in Fig.~\ref{fig: setup}(a), where the triangular planes lie on the $xy$ plane and each layer is stacked along the $z$ direction. 
Here, we adopt a three-sublattice stacking consisting of layer A, layer B, and layer C in Fig.~\ref{fig: setup}(a), i.e., ABCABC$\cdots$ stacking separated by a distance $c$. 
We suppose that each layer does not possess threefold-rotational symmetry, while the system is invariant under the threefold screw operation. 
For simplicity, we take the effect of the threefold-rotational symmetry breaking as the different magnitudes of the exchange interactions rather than the different lattice constant by setting $a=a'$ [Fig.~\ref{fig: setup}(a)]. 
Hereafter, we set $a=a'=c=1$. 

The minimum spin model while keeping threefold screw symmetry but without threefold rotational symmetry in each layer is given by 
\begin{align}
\label{eq: Ham}
\mathcal{H}=&\sum_{\eta}\mathcal{H}^{\perp}_{\eta}+\mathcal{H}^{\parallel}+\mathcal{H}^{{\rm Z}}, \\
\label{eq: Ham_perp}
\mathcal{H}^{\perp}_{\eta}=&  -\sum_{i,j} \left[ J^{(\eta)}_{ij} \bm{S}_{i} \cdot \bm{S}_{j}+ \bm{D}_{ij}^{(\eta)} \cdot  (\bm{S}_{i} \times \bm{S}_{j}) \right], \\
\label{eq: Ham_parallel}
\mathcal{H}^{\parallel}=& -J_{\parallel} \sum_{i,\delta=\pm1} \bm{S}_i \cdot \bm{S}_{i+\delta\hat{z}},\\
\label{eq: Ham_Zeeman}
\mathcal{H}^{{\rm Z}}=&-H \sum_i S_i^z, 
\end{align}
where $\bm{S}_i=(S_i^x, S_i^y, S_i^z)$ represents the classical spin at site $i$. 
The total Hamiltonian $\mathcal{H}$ is divided into the intralayer Hamiltonian $\mathcal{H}^{\perp}_{\eta}$ for $\eta={\rm A}, {\rm B}$, and C, the interlayer Hamiltonian $\mathcal{H}^{\parallel}$, and the Zeeman Hamiltonian $\mathcal{H}^{{\rm Z}}$. 

The intralayer Hamiltonian $\mathcal{H}^{\perp}_{\eta}$ in Eq.~(\ref{eq: Ham_perp}) includes the layer-dependent exchange interaction $J^{(\eta)}_{ij}$ and DM interaction $\bm{D}^{(\eta)}_{ij}$, the latter of which arises from the spin-orbit coupling without the inversion center at the bond.
We consider the polar-type DM vector in the $xy$ plane, whose directions are perpendicular to the inplane nearest-neighbor bond. 
To describe the situation where threefold screw symmetry is preserved while threefold rotational symmetry is broken, we take two coupling constants, $J$ and $J'$, in the intralayer interaction; 
the exchange interactions along the $x$ bond are described by $J'$ and the others are by $J$ for layer A, and $J$ and $J'$ for layers B and C are allocated so that threefold screw symmetry is kept but threefold rotational symmetry is lost, as shown in Fig.~\ref{fig: setup}(a). 
Similarly, the magnitudes of the DM interaction are differently taken as $D$ and $D'$ depending on the bond direction and the layer. 

The interlayer Hamiltonian $\mathcal{H}^{\parallel}$ in Eq.~(\ref{eq: Ham_parallel}) represents the exchange coupling between the different layers. 
We suppose the ferromagnetic coupling constant $J_{\parallel}>0$. 
In addition, we introduce the effect of an external magnetic field in the form of the Zeeman coupling with the magnitude $H$ along the $z$ direction in Eq.~(\ref{eq: Ham_Zeeman}).
In the model in Eq.~(\ref{eq: Ham}), we neglect other anisotropic exchange interactions that arise from the higher-order contributions in terms of the spin-orbit coupling for simplicity.

When setting $J=J'$ and $D=D'$, the model in Eq.~(\ref{eq: Ham}) reduces to the Heisenberg model with the DM interaction, which is invariant under threefold rotational symmetry, i.e., the standard layered triangular lattice.  
This model has been extensively studied as a typical model hosting the SkX in polar and chiral magnets; the SkX to satisfy threefold rotational symmetry is stabilized by the interplay between the intralayer ferromagnetic exchange interaction and the DM interaction in a magnetic field for $J_{\parallel} \geq 0$~\cite{Yi_PhysRevB.80.054416, Mochizuki_PhysRevLett.108.017601,Rowland_PhysRevB.93.020404}.  
Meanwhile, when considering the situation with $J\neq J'$ and $D \neq D'$, the threefold rotational symmetry in each layer is lost, while the threefold screw axis is still present.   
As the SkX usually appears in a threefold-symmetric way, it seems to be destabilized by setting $J\neq J'$ and $D \neq D'$. 
We study the possibility that the SkX is stabilized by focusing on the role of the threefold screw symmetry rather than the threefold rotational one. 

In order to investigate the low-temperature phase diagram in the model in Eq.~(\ref{eq: Ham}), we simplify the intralayer Hamiltonian $\mathcal{H}^{\perp}_{\eta}$ for $\eta=$A, B, and C as 
\begin{align}
\label{eq: Ham_perp2}
\tilde{\mathcal{H}}^{\perp}_\eta=& - \sum_{\nu} \Big[ J^{(\eta)}_\nu \bm{S}^{(\eta)}_{\bm{Q}_{\nu}} \cdot \bm{S}^{(\eta)}_{-\bm{Q}_{\nu}}+ i   \bm{D}^{(\eta)}_\nu \cdot ( \bm{S}^{(\eta)}_{\bm{Q}_{\nu}} \times \bm{S}^{(\eta)}_{-\bm{Q}_{\nu}}) \Big],  
\end{align}
where $\bm{S}^{(\eta)}_{\bm{Q}_{\nu}}$ is the Fourier transform of $\bm{S}_i$ with wave vector $\bm{Q}_\nu$ for the layer $\eta$. 
Here, we take into account the interactions in momentum ($\bm{q}$) space by supposing global minima at $\bm{Q}_1=(\pi/3,0)$, $\bm{Q}_2=(-\pi/6,\sqrt{3}\pi/6)$, $\bm{Q}_3=(-\pi/6,-\sqrt{3}\pi/6)$, $\bm{Q}_4=-\bm{Q}_1$, $\bm{Q}_5=-\bm{Q}_2$, and $\bm{Q}_6=-\bm{Q}_3$ in the interaction tensor $X^{\alpha\beta (\eta)}(\bm{q})$ ($\alpha,\beta=x,y,z$), which is obtained by performing the Fourier transformation of $\mathcal{H}^{\perp}_{\eta}$ in Eq.~(\ref{eq: Ham_perp}), i.e., $\sum_{\bm{q},\eta} [J^{(\eta)}_{\bm{q}} \bm{S}^{(\eta)}_{\bm{q}} \cdot \bm{S}^{(\eta)}_{-\bm{q}}+i   \bm{D}^{(\eta)}_{\bm{q}} \cdot ( \bm{S}^{(\eta)}_{\bm{q}} \times \bm{S}^{(\eta)}_{-\bm{q}})]=\sum_{\bm{q}\eta\alpha\beta} S^{\alpha(\eta)}_{\bm{q}}X^{\alpha\beta (\eta)}(\bm{q}) S^{\beta(\eta)}_{-\bm{q}}$. 
We neglect the contributions from the other $\bm{q}$ components in the interactions for simplicity. 
This assumption is justified when the low-temperature spin configurations are considered and the effect of the interaction at the higher-harmonic ordering vectors is ignored compared to that at $\bm{Q}_\nu$~\cite{leonov2015multiply, Hayami_PhysRevB.103.224418, Hayami_PhysRevB.105.014408,hayami2022multiple}. 
From the threefold screw symmetry, we set the interactions as $J_{\bm{Q}} \equiv J^{(\rm A)}_{\bm{Q}_{1}}=J^{(\rm B)}_{\bm{Q}_{2}}=J^{(\rm C)}_{\bm{Q}_{3}}$, $J'_{\bm{Q}} \equiv J^{(\rm A)}_{\bm{Q}_{2,3}}=J^{(\rm B)}_{\bm{Q}_{1,3}}=J^{(\rm C)}_{\bm{Q}_{1,2}}$, $|\bm{D}_{\bm{Q}}|=D_{\bm{Q}} \equiv |\bm{D}^{(\rm A)}_{\bm{Q}_{1}}|= |\bm{D}^{(\rm B)}_{\bm{Q}_{2}}|= |\bm{D}^{(\rm C)}_{\bm{Q}_{3}}|$, and $|\bm{D}'_{\bm{Q}}|=D'_{\bm{Q}} \equiv |\bm{D}^{(\rm A)}_{\bm{Q}_{2,3}}|= |\bm{D}^{(\rm B)}_{\bm{Q}_{1,3}}|= |\bm{D}^{(\rm C)}_{\bm{Q}_{1,2}}|$; 
the interaction at $\bm{Q}_1$ for layer A is equivalent with those at $\bm{Q}_2$ for layer B and at $\bm{Q}_3$ for layer C, as shown in Fig.~\ref{fig: setup}(b). 

Then, we examine the total Hamiltonian written by 
\begin{align}
\label{eq: Ham2}
\mathcal{H}=&\sum_{\eta}\tilde{\mathcal{H}}^{\perp}_{\eta}+\mathcal{H}^{\parallel}+\mathcal{H}^{{\rm Z}}. 
\end{align}
In the following, we set $J_{\bm{Q}}=1$ as the energy unit of the model in Eq.~(\ref{eq: Ham2}), and we fix $D_{\bm{Q}}=0.2$~\cite{comment_DM}. 
In addition, we introduce the parameter to measure the degree of the threefold symmetry breaking as $\kappa=J'_{\bm{Q}}/J_{\bm{Q}}=D'_{\bm{Q}}/D_{\bm{Q}}$, where $\kappa=1$ stands for the situation in the presence of the threefold rotational symmetry; the SkX is stabilized under the magnetic field, where the model in Eq.~(\ref{eq: Ham2}) reduces to the layered triangular-lattice model with the uniform DM interaction~\cite{Hayami_PhysRevB.105.014408}.   
We investigate the magnetic phase diagram while changing $\kappa$, $J_{\parallel}$, and $H$. 
Specifically, we show the phase diagram in the plane of $J_{\parallel}$ and $H$ in Fig.~\ref{fig: PD_JFdep} in Sec.~\ref{sec: Instability toward skyrmion crystal phase} and that in the plane of $\kappa$ and $H$ in Fig.~\ref{fig: PD_ratiodep} in Sec.~\ref{sec: Skyrmion crystal phases with fractional skyrmion numbers}. 
These phase diagrams in the wide range of the model parameters will be a reference for studies based on the ab initio calculations once the SkX-hosting materials are discovered under the space groups with the screw axis like $P3_1$ (\# 144), $P3_2$ (\# 145), $P3_1 12$ (\# 151), $P3_1 21$ (\# 152), $P3_2 12$ (\# 153), and , $P3_2 21$ (\# 154) in experiments.

The magnetic phase diagram of the layered spin model in Eq.~(\ref{eq: Ham2}) is constructed by performing the simulated annealing. 
The simulations have been performed for a system size with $N=3\times L^2$ and $L=48$ under the periodic boundary conditions in all the directions. 
The procedures of the simulations are as follows. 
First, we start from a random spin configuration at a high temperature, which is typically selected as $T_0/J=1$-$10$ where we set the Boltzmann constant $k_{\rm B}$ as unity. 
Then we reduce the temperature with a rate $T_{n+1}/J=\alpha T_n/J$, where $T_n/J$ is the $n$th-step temperature and $\alpha=0.99999$-$0.999999$, down to the final temperature $T/J=0.001$. 
We choose large $T_0$ and large $\alpha$ when analyzing the vicinity of the phase boundaries.
At each temperature, we perform the standard Metropolis local updates in real space. 
At the final temperature, we perform $10^5$-$10^6$ Monte Carlo sweeps for measurements. 
We also start the simulations from the spin patterns obtained at low temperatures in the vicinity of the phase boundaries between different magnetic phases.

For later convenience, we introduce the spin- and chirality-related quantities, which are used for the identification of the magnetic phases obtained by the simulated annealing. 
The spin structure factor and $\bm{q}$ component of magnetic moments for layer $\eta$ are represented by 
\begin{align}
S_{\eta}^\alpha(\bm{q})&= \frac{1}{L^2} \sum_{i,j \in \eta} S^{\alpha}_i S^{\alpha}_j e^{i\bm{q}\cdot (\bm{r}_i-\bm{r}_j)},  \\
m^\alpha_{\eta\bm{q}}&=\sqrt{\frac{S^\alpha_\eta(\bm{q})}{L^2}}, 
\end{align}
for $\alpha=x,y,z$. 
The site indices $i$ and $j$ are taken for the layer $\eta=$A, B, and C. 
We also compute $S_{\eta}^{xy}(\bm{q})=S_{\eta}^x(\bm{q})+S_{\eta}^y(\bm{q})$ and $m^{xy}_{\eta\bm{q}}=\sqrt{S^{xy}_\eta(\bm{q})/L^2}$.
The net magnetization for each layer is given by $M^\alpha_{\eta}=(1/L^2)\sum_{i \in \eta}S^{\alpha}_{i}$.

The spin scalar chirality is represented by 
\begin{align}
\chi^{\rm sc}_{\eta} = \frac{1}{L^2} \sum_{\bm{R}\in \eta} \bm{S}_{i} \cdot (\bm{S}_j \times \bm{S}_k).
\end{align}
Here, $\bm{R}$ represents the position vector at the centers of triangles, where the sites $i$, $j$, and $k$ are the triangle vertices at $\bm{R}$ in the counterclockwise order. 
The magnetic ordering with nonzero $\chi^{\rm sc}=\chi^{\rm sc}_{\rm A}+ \chi^{\rm sc}_{\rm B}+\chi^{\rm sc}_{\rm C}$ exhibits the topological (intrinsic) Hall effect owing to an emergent magnetic field.

\section{Instability toward skyrmion crystal phase}
\label{sec: Instability toward skyrmion crystal phase}

\begin{figure}[t!]
\begin{center}
\includegraphics[width=1.0 \hsize ]{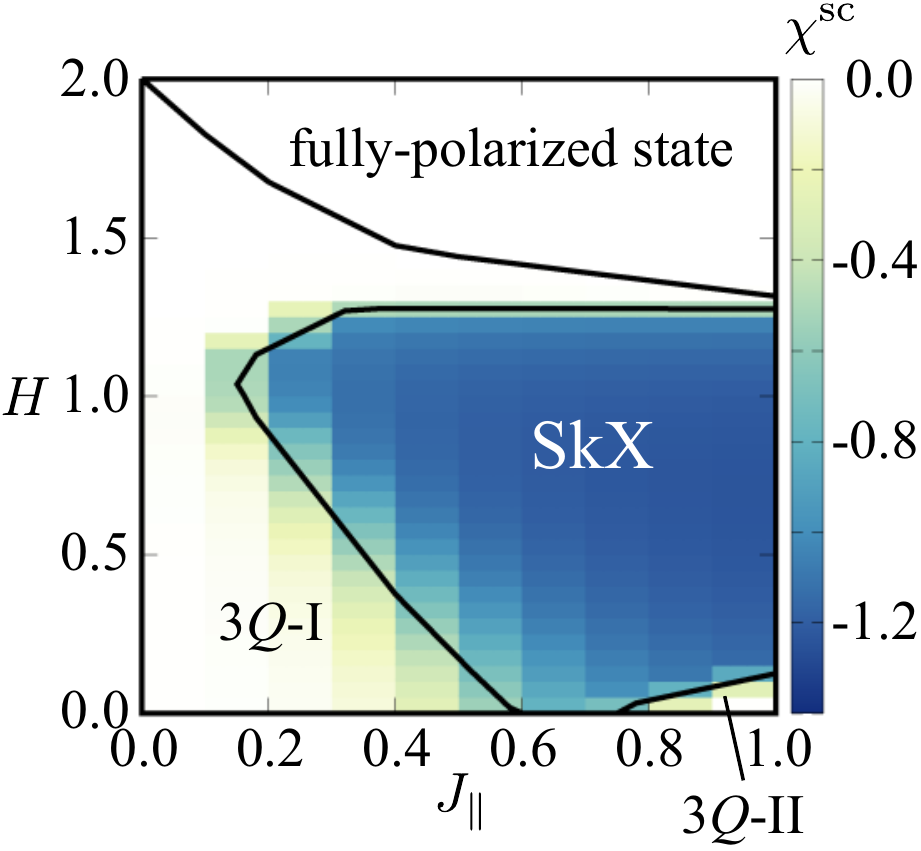} 
\caption{
\label{fig: PD_JFdep}
Phase diagram of the model in Eq.~(\ref{eq: Ham2}) obtained by the simulated annealing at $\kappa=0.4$ while changing $J_{\parallel}$ and $H$. 
The contour shows the scalar chirality $\chi^{\rm sc}$. 
}
\end{center}
\end{figure}

We first show the magnetic phase diagram while varying $J_{\parallel}$ and $H$ for fixed $\kappa=0.4$. 
Figure~\ref{fig: PD_JFdep} shows the result obtained by the simulated annealing, where the color map shows the scalar chirality $\chi^{\rm sc}$. 
There are three magnetic phases in addition to the fully-polarized state with $\bm{S}_i\simeq (0,0,1)$ for large $H$. 
Despite the absence of the threefold rotational symmetry in the system, the SkX, which is identified as the state with the integer skyrmion number at $-1$, appears in a wide range of parameters. 
In particular, one finds that the SkX is stabilized even without the magnetic field for $0.6 \lesssim J_{\parallel}\lesssim 0.75$. 
In the following, we discuss the behavior of spin- and chirality-related quantities for the weak interlayer coupling in Sec.~\ref{sec: Weak interlayer coupling}, the intermediate interlayer coupling in Sec.~\ref{sec: Intermediate interlayer coupling}, and the strong interlayer coupling in Sec.~\ref{sec: Strong interlayer coupling}. 
Although we here focus on the low-temperature spin and chirality configurations in each phase in Fig.~\ref{fig: PD_JFdep}, the obtained phases remain stable when considering the effect of thermal fluctuations at finite temperatures~\cite{hayami2021phase,Hayami_10.1088/1367-2630/ac3683,kato2022magnetic}.

\subsection{Weak interlayer coupling}
\label{sec: Weak interlayer coupling}

\begin{figure}[t!]
\begin{center}
\includegraphics[width=1.0 \hsize ]{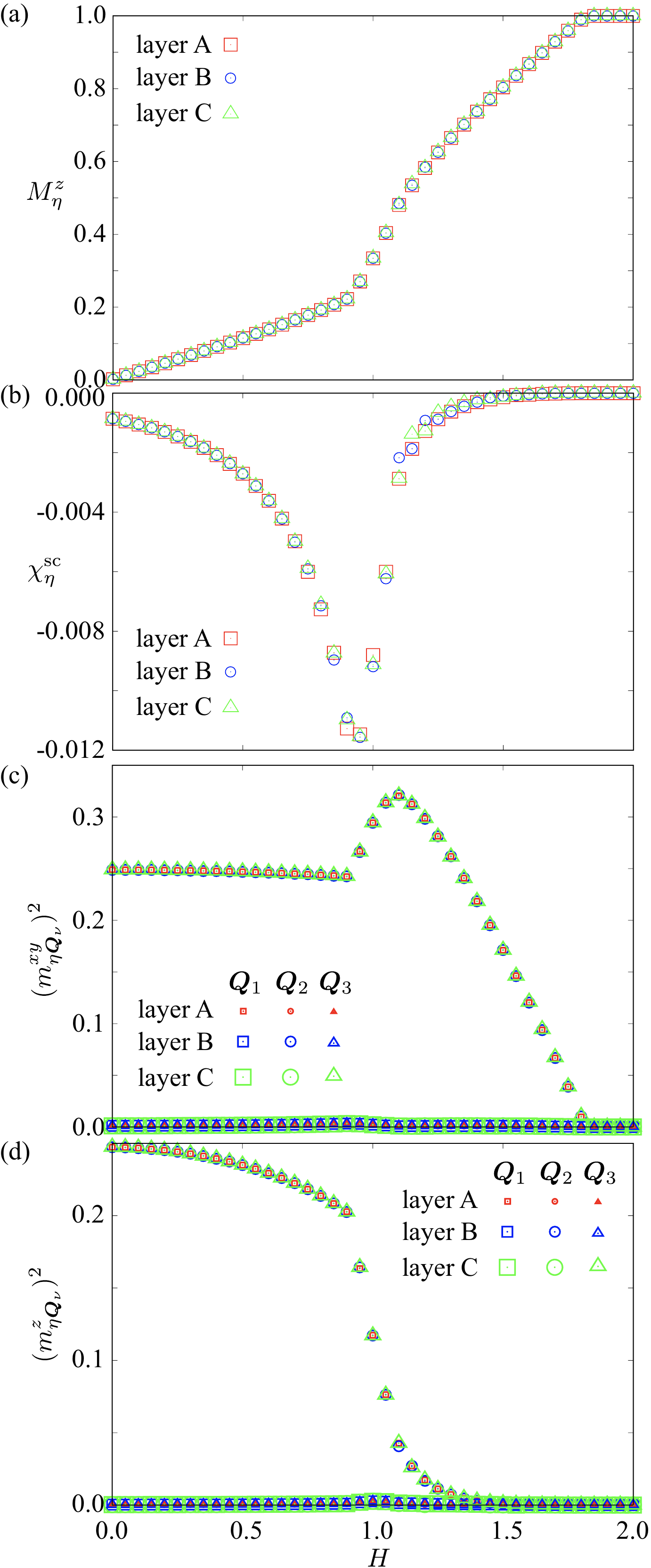} 
\caption{
\label{fig: mq_J=0.1}
$H$ dependences of (a) $M^z_\eta$, (b) $\chi^{\rm sc}_\eta$, (c) $(m^{xy}_{\eta \bm{Q}_\nu})^2$, and (d) $(m^{z}_{\eta \bm{Q}_\nu})^2$ for $J_{\parallel}=0.1$. 
}
\end{center}
\end{figure}

\begin{figure*}[t!]
\begin{center}
\includegraphics[width=0.85 \hsize ]{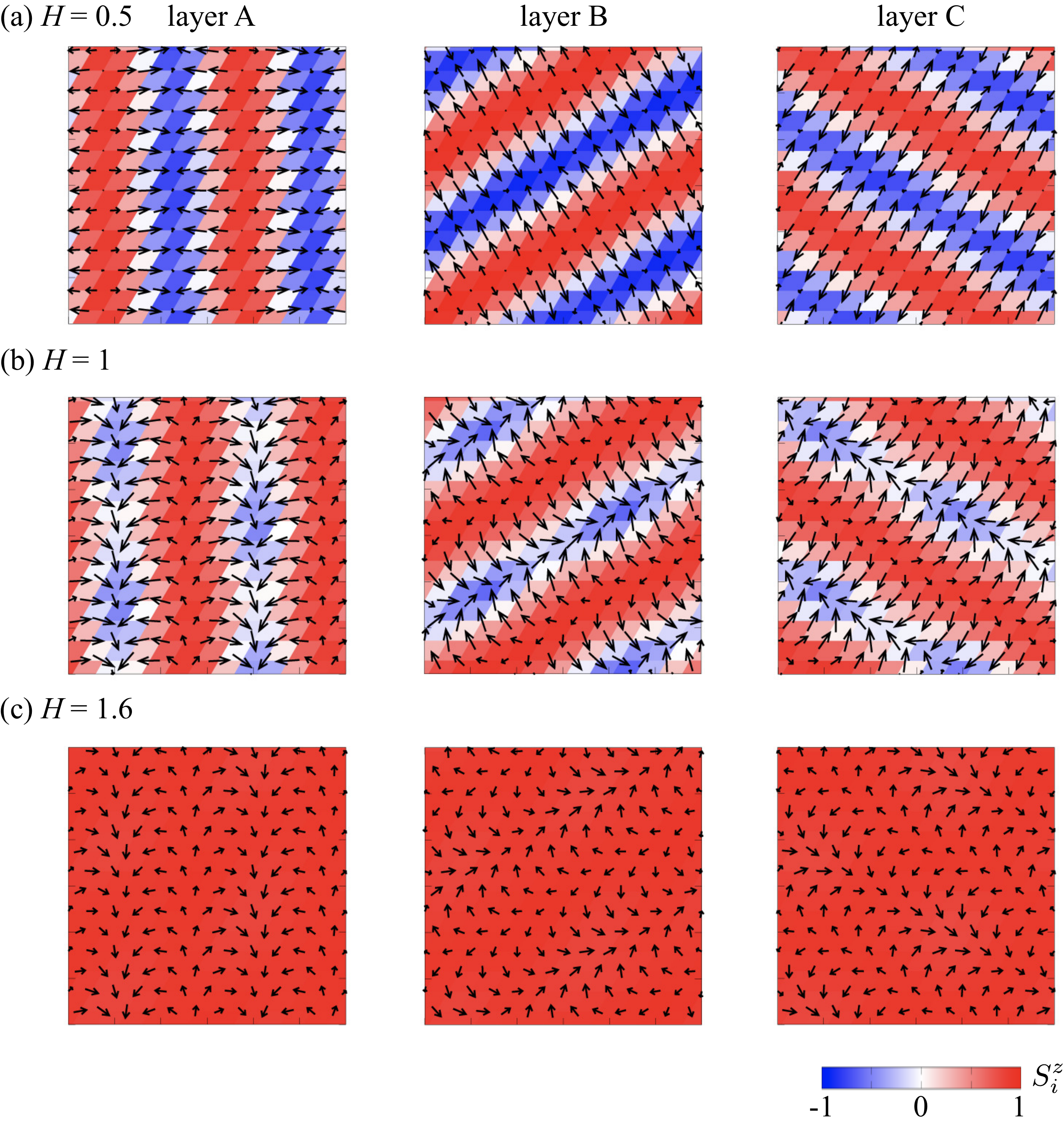} 
\caption{
\label{fig: spin_J=0.1}
Real-space spin configurations of the 3$Q$-I state on the layers A (left), B (middle), and C (right) at (a) $H=0.5$, (b) $H=1$, and (c) $H=1.6$ for $J_{\parallel}=0.1$. 
The color represents the $z$ component of the spin moment, and the arrows stand for the $xy$ components. 
}
\end{center}
\end{figure*}

For $J_{\parallel}=0$, the layers are completely decoupled, and the optimal spin configurations in each layer are determined by the intralayer interaction. 
When considering layer A, the single-$Q$ spiral state at $\bm{Q}_1$ appears. 
The spiral plane lies on the $xz$ plane for $H \lesssim 0.95$ (vertical spiral), while it continuously changes to the $xy$ plane (conical spiral) while increasing $H$. 
Such a change of the spiral plane against $H$ is common to the spin model with the exchange interaction and the DM interaction, e.g., the Heisenberg model in polar and chiral magnets~\cite{Buhrandt_PhysRevB.88.195137}. 
Similarly, the same behavior is found in layers B and C, where the single-$Q$ spiral states at $\bm{Q}_2$ and $\bm{Q}_3$ are stabilized, respectively. 
Thus, the Bragg peaks appear at $\bm{Q}_1$, $\bm{Q}_2$, and $\bm{Q}_3$ in the whole system despite the single-$Q$ spin configuration in each layer, so we call this state a 3$Q$-I state. 

The introduction of $J_{\parallel}$ modulates the single-$Q$ spin configuration in each layer. 
We show the uniform magnetization $M^z_\eta$ in Fig.~\ref{fig: mq_J=0.1}(a), the scalar chirality $\chi^{\rm sc}_\eta$ in Fig.~\ref{fig: mq_J=0.1}(b), the $xy$ component of magnetic moments $(m^{xy}_{\eta \bm{Q}_\nu})^2$ in Fig.~\ref{fig: mq_J=0.1}(c), and the $z$ component of magnetic moments $(m^{z}_{\eta \bm{Q}_\nu})^2$ in Fig.~\ref{fig: mq_J=0.1}(d) in each layer. 
The behaviors of each quantity are similar to each other except for the intermediate-field region, although the dominant $\bm{Q}_\nu$ component is different in each layer; the spiral waves in layers A, B, and C are characterized by those at the $\bm{Q}_1$, $\bm{Q}_2$, and $\bm{Q}_3$ components, respectively. 
In other words, the spin configurations are invariant under the threefold screw operation~\cite{comment_screw}. 
The effect of the threefold screw symmetry is found in the real-space spin configuration in each layer for several $H$ in Fig.~\ref{fig: spin_J=0.1}. 
From Figs.~\ref{fig: spin_J=0.1}(a)-\ref{fig: spin_J=0.1}(c), one finds that the spiral plane at dominant $\bm{Q}_\eta$ changes from the $xz$ or $yz$ plane to the $xy$ plane, which is similar to the case at $J_{\parallel}=0$. 

The difference from the result at $J_{\parallel}=0$ is that the single-$Q$ spiral spin configuration in each layer is modulated so as to have the amplitudes of $\bm{m}_{\eta\bm{Q}_\nu}$ at the other two $\bm{Q}_\eta$, as shown in Figs.~\ref{fig: mq_J=0.1}(c) and \ref{fig: mq_J=0.1}(d).  
For example, the $\bm{Q}_2$ and $\bm{Q}_3$ components become nonzero in layer A while keeping $m^{xy}_{{\rm A}\bm{Q}_2}=m^{xy}_{{\rm A}\bm{Q}_3}$ and $m^{z}_{{\rm A}\bm{Q}_2}=m^{z}_{{\rm A}\bm{Q}_3}$ for $H \lesssim 1$ and $m^{xy}_{{\rm A}\bm{Q}_2} \neq m^{xy}_{{\rm A}\bm{Q}_3} \neq 0$ and $m^{z}_{{\rm A}\bm{Q}_2} \neq m^{z}_{{\rm A}\bm{Q}_3} \neq 0$ for $H \gtrsim 1$, although their values are much smaller than those at $\bm{Q}_{\eta}$.  
The appearance of the triple-$Q$ modulation is also found in the real-space spin configuration especially for $H=1$ in Fig.~\ref{fig: mq_J=0.1}(b). 
Accordingly, the scalar chirality $\chi^{\rm sc}_\eta$ is slightly induced in all the regions in each layer. 
Especially, its magnitude takes the maximum value around $H \simeq 1$, where the spiral plane is changed from the $xz$ or $yz$ plane to the $xy$ plane, as shown in Fig.~\ref{fig: mq_J=0.1}(b). 
It is noted that the skyrmion number is zero in this spin configuration.

\subsection{Intermediate interlayer coupling}
\label{sec: Intermediate interlayer coupling}

\begin{figure}[t!]
\begin{center}
\includegraphics[width=1.0 \hsize ]{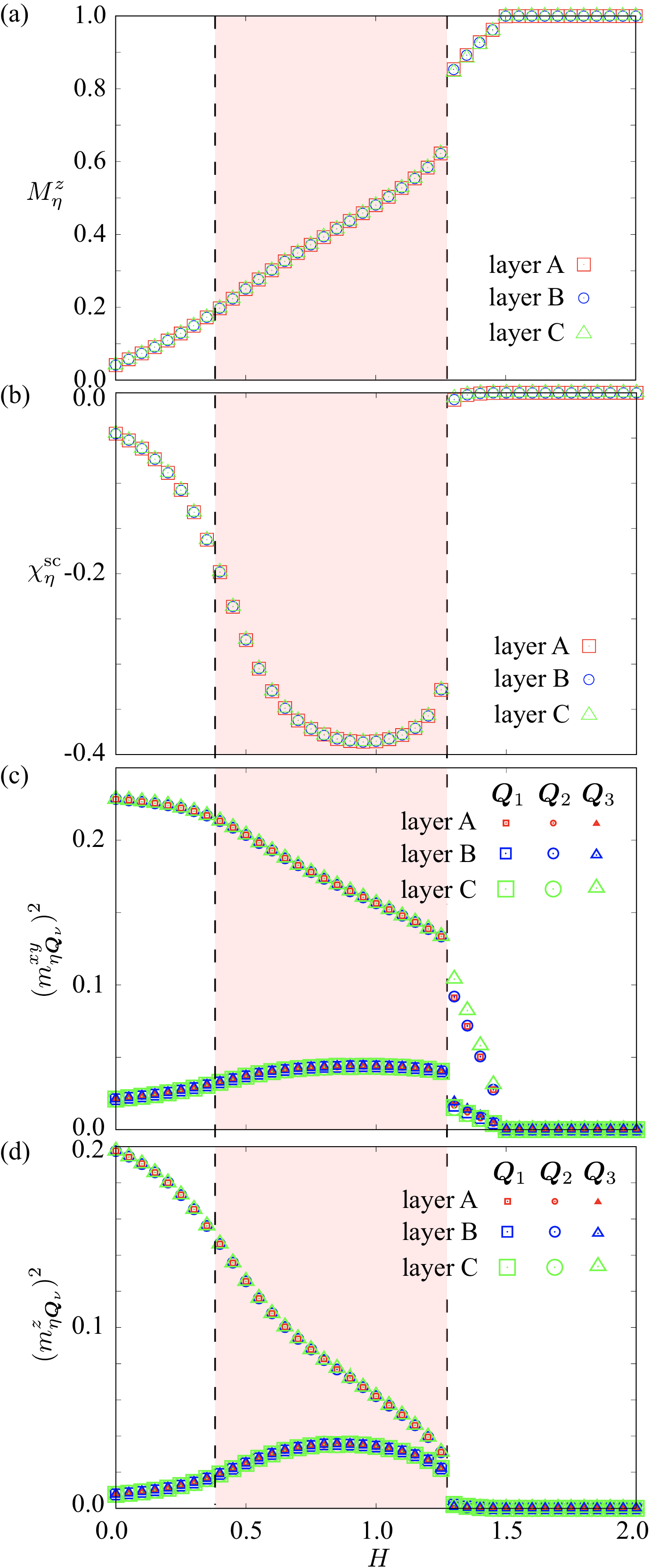} 
\caption{
\label{fig: mq_J=0.4}
$H$ dependences of (a) $M^z_\eta$, (b) $\chi^{\rm sc}_\eta$, (c) $(m^{xy}_{\eta \bm{Q}_\nu})^2$, and (d) $(m^{z}_{\eta \bm{Q}_\nu})^2$ for $J_{\parallel}=0.4$. 
The vertical dashed lines represent the phase boundaries between the SkX and the other states.
}
\end{center}
\end{figure}

\begin{figure*}[t!]
\begin{center}
\includegraphics[width=0.85 \hsize ]{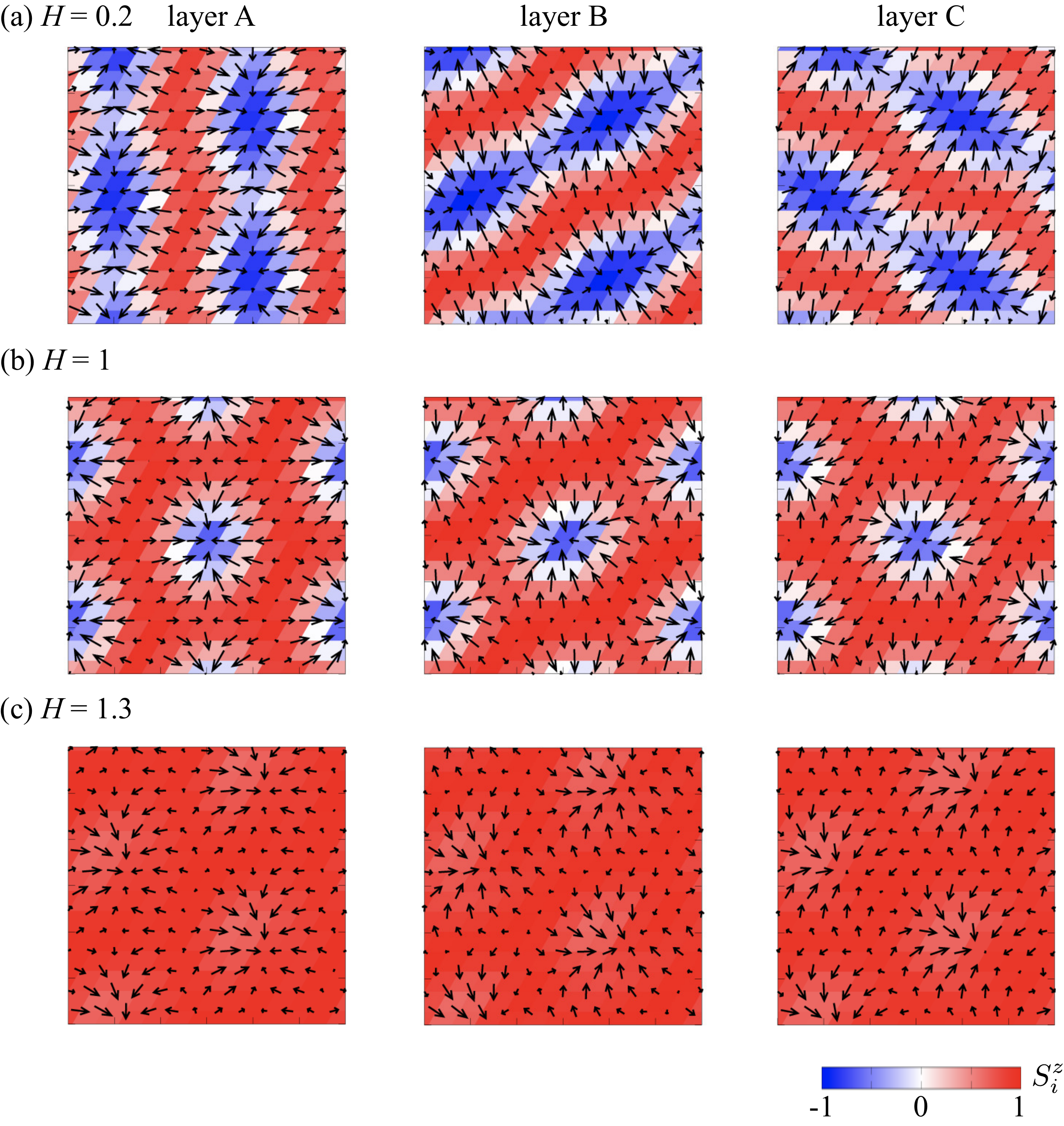} 
\caption{
\label{fig: spin_J=0.4}
Real-space spin configurations of (a) the 3$Q$-I state at $H=0.2$, (b) the SkX at $H=1$, and (c) the 3$Q$-I state at $H=1.3$ on the layers A (left), B (middle), and C (right) for $J_{\parallel}=0.4$. 
The color represents the $z$ component of the spin moment, and the arrows stand for the $xy$ components. 
}
\end{center}
\end{figure*}

\begin{figure}[t!]
\begin{center}
\includegraphics[width=1.0 \hsize ]{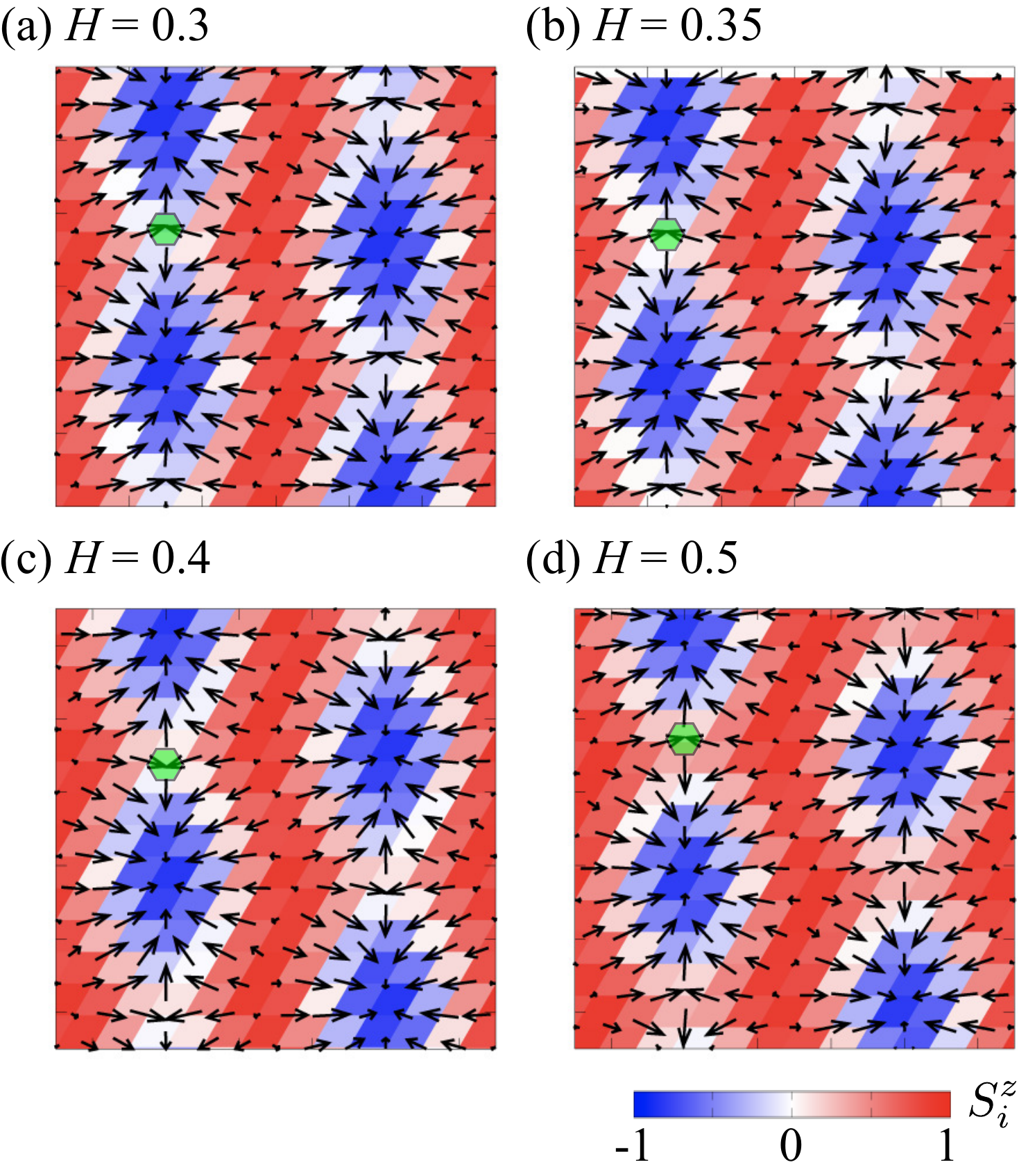} 
\caption{
\label{fig: spin_SkX}
Real-space spin configurations of (a,b) the 3$Q$-I state at (a) $H=0.3$ and (b) $H=0.35$ and (c,d) the SkX at (c) $H=0.4$ and (d) $H=0.5$ on layer A for $J_{\parallel}=0.4$. 
The color represents the $z$ component of the spin moment, and the arrows stand for the $xy$ components. 
}
\end{center}
\end{figure}

Figure~\ref{fig: mq_J=0.4} shows the results for the intermediate interlayer coupling $J_{\parallel}=0.4$, which corresponds to those in Fig.~\ref{fig: mq_J=0.1}. 
Although the magnetization becomes nonzero for $H=0$ in contrast to the weak interlayer coupling [Fig.~\ref{fig: mq_J=0.4}(a)], the behaviors of other quantities for small $H$ are similar to those in the 3$Q$-I state in Fig.~\ref{fig: mq_J=0.1}; each layer has the anisotropic triple-$Q$ components of $(m^{xy}_{\eta \bm{Q}_\nu})^2$ and $(m^{z}_{\eta \bm{Q}_\nu})^2$, e.g., $(m^{xy}_{{\rm A} \bm{Q}_1})^2 > (m^{xy}_{{\rm A} \bm{Q}_2})^2=(m^{xy}_{{\rm A} \bm{Q}_3})^2$ and $(m^{z}_{{\rm A} \bm{Q}_1})^2 > (m^{z}_{{\rm A} \bm{Q}_2})^2=(m^{z}_{{\rm A} \bm{Q}_3})^2$ for layer A, as shown in Figs.~\ref{fig: mq_J=0.4}(c) and \ref{fig: mq_J=0.4}(d). 
The magnitudes of $(m^{xy}_{\eta \bm{Q}_\nu})^2$ and $(m^{z}_{\eta \bm{Q}_\nu})^2$ at the subdominant peaks like $(m^{xy}_{{\rm A} \bm{Q}_2})^2, (m^{xy}_{{\rm A} \bm{Q}_3})^2$ and $(m^{z}_{{\rm A} \bm{Q}_2})^2, (m^{z}_{{\rm A} \bm{Q}_3})^2$ become larger compared to those for the weak interlayer coupling, as shown in Figs.~\ref{fig: mq_J=0.1}(c), \ref{fig: mq_J=0.1}(d), \ref{fig: mq_J=0.4}(c), and \ref{fig: mq_J=0.4}(d). 
The enhancement of magnetic moments at the subdominant wave vectors leads to a further modulation from the stripe structure to the hexagonal structure, as shown by the real-space spin configuration in Fig.~\ref{fig: spin_J=0.4}(a). 
Furthermore, the scalar chirality $\chi^{\rm sc}_{\eta}$ becomes larger as $J_{\parallel}$ increases owing to the development of magnetic moments at the subdominant wave vectors, as shown in Fig.~\ref{fig: mq_J=0.4}(b). 

While increasing $H$, $\chi^{\rm sc}_{\eta}$ is enhanced as well as $M^z_\eta$ shown in Figs.~\ref{fig: mq_J=0.4}(a) and \ref{fig: mq_J=0.4}(b), and then the skyrmion number becomes $-1$ in each layer for $H \gtrsim 0.4$ without any jumps in spin and chirality quantities. 
The quantized skyrmion number means the emergence of the SkX. 
Indeed, the real-space spin configuration clearly exhibits the periodic alignment of the skyrmion spin textures in each layer, as shown in Fig.~\ref{fig: spin_J=0.4}(b). 
It is noted that the skyrmion cores denoted as $S_i^z = -1$ are located at the interstitial site~\cite{Hayami_PhysRevResearch.3.043158} and are elongated along the direction perpendicular to dominant $\bm{Q}_{\eta}$ in each layer. 

This result indicates that the SkX can appear in the systems with the threefold screw axis when the layer interaction becomes relatively large; $J_{\parallel} \gtrsim 0.15$ is enough to stabilize the SkX at $\kappa=0.4$. 
As will be discussed in Fig.~\ref{fig: PD_ratiodep}, the critical value of $J_{\parallel}$ depends on $\kappa$ which shows the degree of the anisotropic interaction in the triangular-lattice structure. 
For example, the SkX at $J_{\parallel}=0.2$ and $\kappa=0.4$ is destabilized while decreasing $\kappa$. 
Thus, the appearance of the SkX is due to the interplay between $J_{\parallel}$ and $\kappa$. 
Moreover, the region where the SkX is stabilized is extended while increasing $J_{\parallel}$, and it stabilizes even at a zero field for $0.6 \lesssim J_{\parallel}\lesssim 0.75$. 
The extension of the SkX to the low-field region against $J_{\parallel}$ is reasonable since an effective coupling in the form of $(\bm{S}_{\bm{0}}\cdot \bm{S}_{\bm{Q}_1})(\bm{S}_{\bm{Q}_2}\cdot \bm{S}_{\bm{Q}_3})$ in the SkX becomes stronger while increasing $J_{\parallel}$. 

The continuous phase transition from the 3$Q$-I state to the SkX found at $J_{\parallel}=0.4$ in Fig.~\ref{fig: mq_J=0.4} is understood from the real-space spin configurations. 
We show the real-space spin configurations for layer A in the vicinity of the phase boundary between the 3$Q$-I and SkX phases in Fig.~\ref{fig: spin_SkX}. 
According to the development of $M^z_{\rm A}$, $(m^{xy}_{{\rm A} \bm{Q}_2})^2$, $(m^{xy}_{{\rm A} \bm{Q}_3})^2$, $(m^{z}_{{\rm A} \bm{Q}_2})^2$, and $(m^{z}_{{\rm A} \bm{Q}_3})^2$, the sign of $S_i^z$ at the position denoted by green hexagons in Fig.~\ref{fig: spin_SkX} changes with an increase of $H$ while keeping the inplane spin textures; $S_i^z<0$ in the 3$Q$-I state for $H=0.3$ [Fig.~\ref{fig: spin_SkX}(a)] and $H=0.35$ [Fig.~\ref{fig: spin_SkX}(b)], while $S_i^z>0$ in the SkX for $H=0.4$ [Fig.~\ref{fig: spin_SkX}(c)] and $H=0.5$ [Fig.~\ref{fig: spin_SkX}(d)]. 
The sign change of $S_i^z$ leads to the sign reversal of local scalar chirality, which results in the emergent SkX for larger $H$. 
In other words, the smooth change of $S_i^z$ owing to an increase of $H$ makes the continuous phase transition with different skyrmion numbers possible. 

In a high-field region, the SkX changes into the 3$Q$-I state, where $\chi_\eta^{\rm sc}$ and $(m^z_{\eta \bm{Q}_\nu})^2$ are largely suppressed shown in Figs.~\ref{fig: mq_J=0.4}(b) and \ref{fig: mq_J=0.4}(d), and accordingly, the skyrmion number becomes zero.
The real-space spin configuration obtained by the simulated annealing at $H=1.3$ is shown in Fig.~\ref{fig: spin_J=0.4}(c). 
Similar to Fig.~\ref{fig: spin_J=0.4}(a), the dominant modulation is described by $(m^{xy}_{{\rm A} \bm{Q}_1})^2$, $(m^{xy}_{{\rm B} \bm{Q}_2})^2$, and $(m^{xy}_{{\rm C} \bm{Q}_3})^2$ for layers A, B, and C, respectively, although the intensities between $(m^{xy}_{{\rm A,B} \bm{Q}_1})^2$ and $(m^{xy}_{{\rm C} \bm{Q}_2})^2$ are slightly different, as shown in Fig.~\ref{fig: mq_J=0.4}(c). 
This state is similar to a triple-$Q$ vortex crystal found in frustrated magnets~\cite{Kamiya_PhysRevX.4.011023, Hayami_PhysRevB.94.174420} and itinerant magnets~\cite{Hayami_PhysRevB.103.054422,yambe2021skyrmion}.

\begin{figure}[t!]
\begin{center}
\includegraphics[width=1.0 \hsize ]{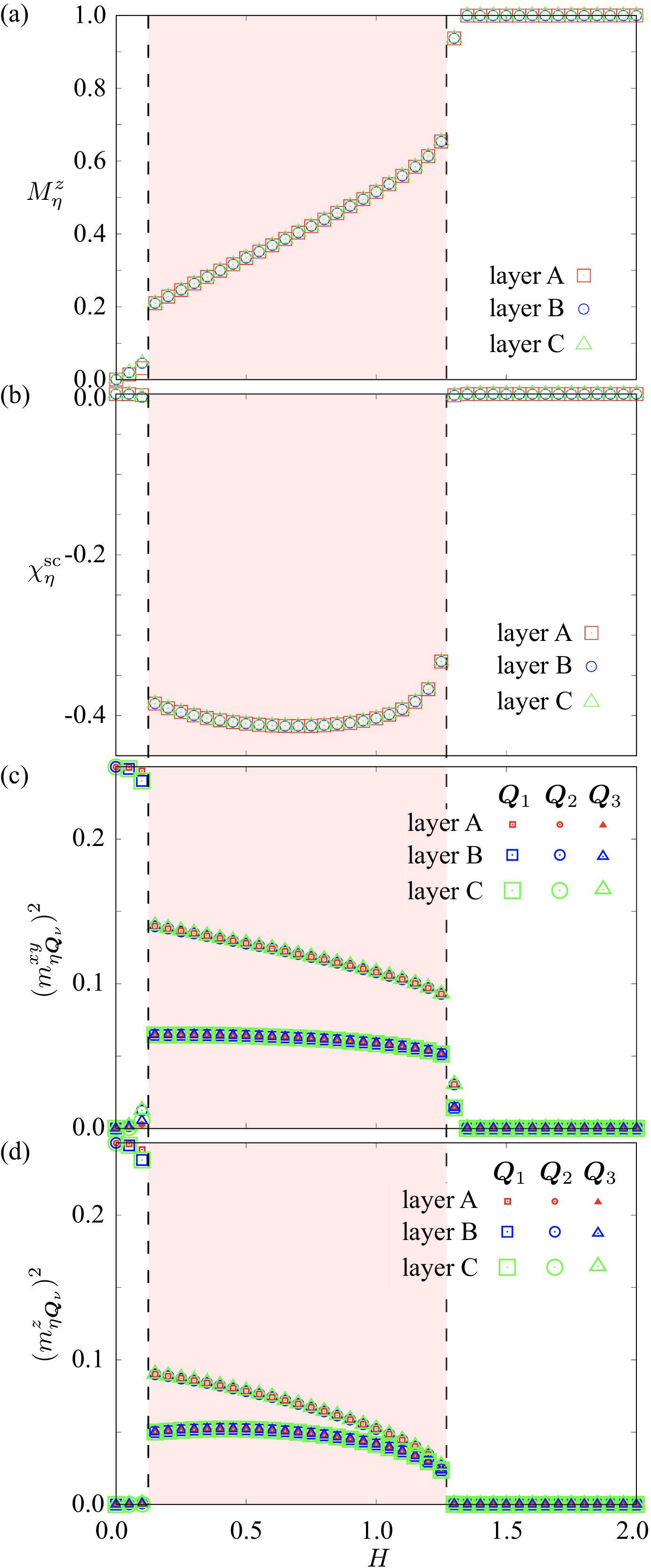} 
\caption{
\label{fig: mq_J=1.0}
$H$ dependences of (a) $M^z_\eta$, (b) $\chi^{\rm sc}_\eta$, (c) $(m^{xy}_{\eta \bm{Q}_\nu})^2$, and (d) $(m^{z}_{\eta \bm{Q}_\nu})^2$ for $J_{\parallel}=1$. 
The vertical dashed lines represent the phase boundaries between the SkX and the other states.
}
\end{center}
\end{figure}

\begin{figure*}[t!]
\begin{center}
\includegraphics[width=0.85 \hsize ]{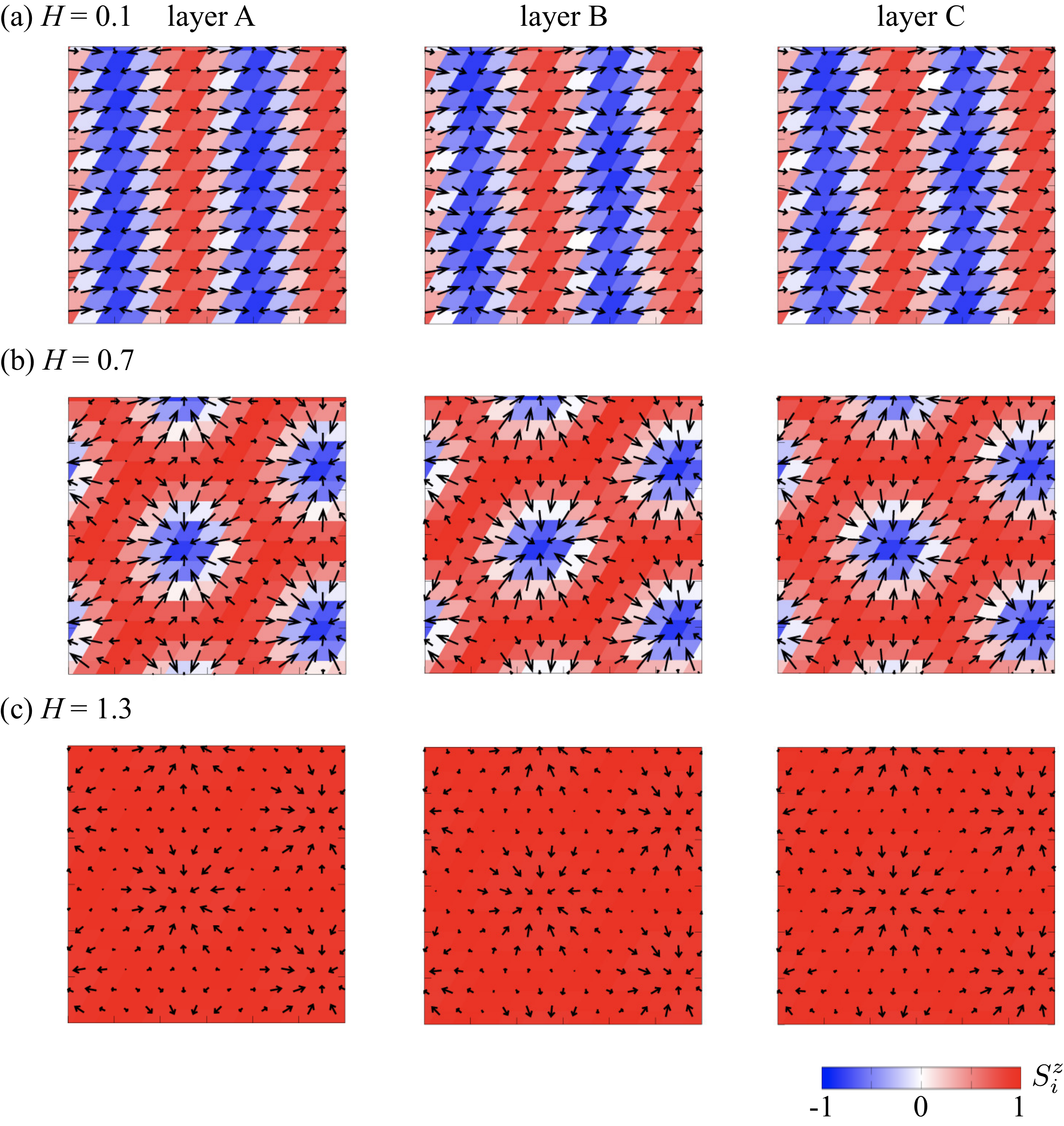} 
\caption{
\label{fig: spin_J=1.0}
Real-space spin configurations of (a) the 3$Q$-II state at $H=0.1$, (b) the SkX at $H=0.7$, and (c) the 3$Q$-I state at $H=1.3$ on the layers A (left), B (middle), and C (right) for $J_{\parallel}=1$. 
The color represents the $z$ component of the spin moment, and the arrows stand for the $xy$ components. 
}
\end{center}
\end{figure*}

\subsection{Strong interlayer coupling}
\label{sec: Strong interlayer coupling}

In the case of strong interlayer coupling for $J_{\parallel}>0.75$, the SkX in the low-field region is replaced with the other triple-$Q$ state denoted as a 3$Q$-II state, as shown in Fig.~\ref{fig: PD_JFdep}. 
In this state, the dominant $\bm{Q}_\nu$ component of magnetic moments is common to layers A, B, and C. 
For example, the spin state is mainly characterized by $m^{xy}_{{\rm A}\bm{Q}_1}$, $m^{xy}_{{\rm B}\bm{Q}_1}$, and $m^{xy}_{{\rm C}\bm{Q}_1}$ [$m^{z}_{{\rm A}\bm{Q}_1}$, $m^{z}_{{\rm B}\bm{Q}_1}$, and $m^{z}_{{\rm C}\bm{Q}_1}$] at $H=0.1$, as shown in the case of $J_{\parallel}=1$ in Fig.~\ref{fig: mq_J=1.0}(c) [Fig.~\ref{fig: mq_J=1.0}(d)]. 
In other words, the spiral direction is the same for all layers A, B, and C, which is clearly found in the real-space spin configuration in Fig.~\ref{fig: spin_J=1.0}(a).  
This is understood from a magnetic frustration in the consideration of $\kappa$ and $J_{\parallel}$;  
the small $\kappa$ or $J_{\parallel}$ tends to favor the spiral modulation along the different direction (3$Q$-I state)depending on the layer so as to gain the energy by $J_{\bm{Q}_\nu}$ and $D_{\bm{Q}_\nu}$ in Eq.~(\ref{eq: Ham_perp2}), whereas the large $\kappa$ or $J_{\parallel}$ tends to favor the spiral modulation along the same direction independent of the layer (3$Q$-II state) so as to gain the energy by $J'_{\bm{Q}_\nu}$, $D'_{\bm{Q}_\nu}$, and $J_{\parallel}$. 
Remarkably, there is an instability toward the SkX in the boundary region between the two triple-$Q$ phases, where the effect of frustration is maximized, as discussed in Sec.~\ref{sec: Intermediate interlayer coupling}. 
The region where the 3$Q$-II state is stabilized becomes larger while increasing $J_{\parallel}$ (Fig.~\ref{fig: PD_JFdep}), which is consistent with the above argument. 

Upon increasing $H$ from the 3$Q$-II state at $J_{\parallel}=1$, the state turns into the SkX at $H \simeq 0.15$, as shown in Fig.~\ref{fig: mq_J=1.0}. 
The behaviors of spin and chirality quantities against $H$ are similar to those in Fig.~\ref{fig: mq_J=0.4}. 
Similar to the case of $J_{\parallel}=0.4$, the SkX turns into the 3$Q$-I state while further increasing $H$. 
The spin configurations of the SkX and the 3$Q$-I state are shown in Figs.~\ref{fig: spin_J=1.0}(b) and \ref{fig: spin_J=1.0}(c), which resembles those in Figs.~\ref{fig: spin_J=0.4}(b) and \ref{fig: spin_J=0.4}(c).

\section{Skyrmion crystals with fractional skyrmion numbers}
\label{sec: Skyrmion crystal phases with fractional skyrmion numbers}

\begin{figure}[t!]
\begin{center}
\includegraphics[width=1.0 \hsize ]{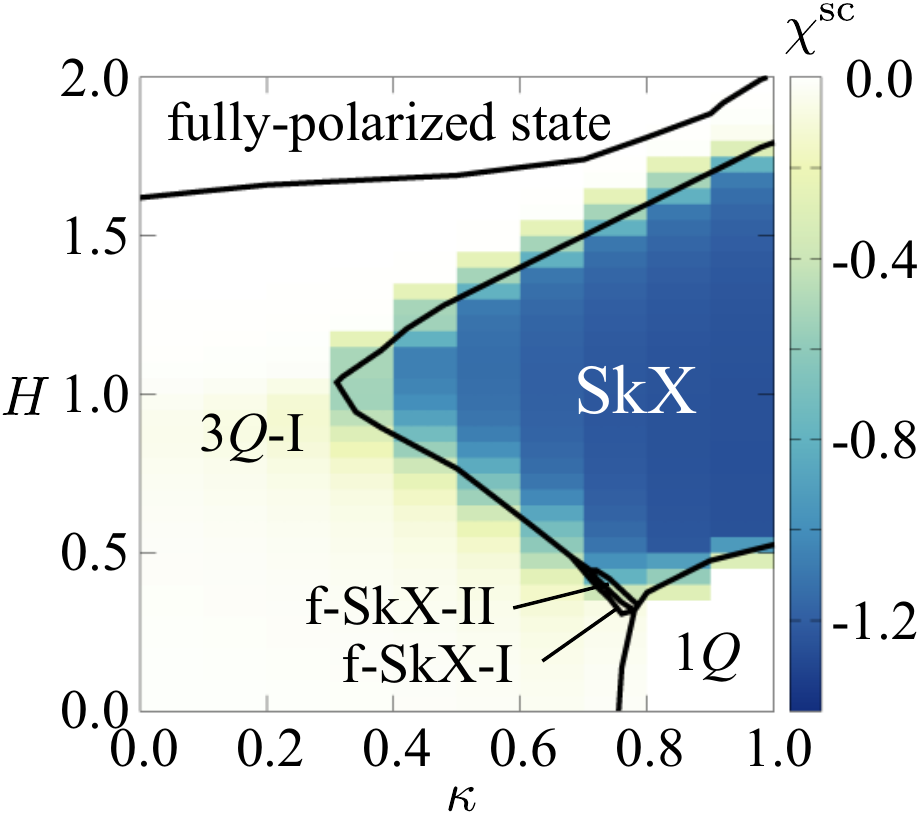} 
\caption{
\label{fig: PD_ratiodep}
Phase diagram of the model in Eq.~(\ref{eq: Ham2}) obtained by the simulated annealing at $J_{\parallel}=0.2$ while changing $\kappa$ and $H$. 
The contour shows the scalar chirality $\chi^{\rm sc}$. 
}
\end{center}
\end{figure}

\begin{figure}[t!]
\begin{center}
\includegraphics[width=1.0 \hsize ]{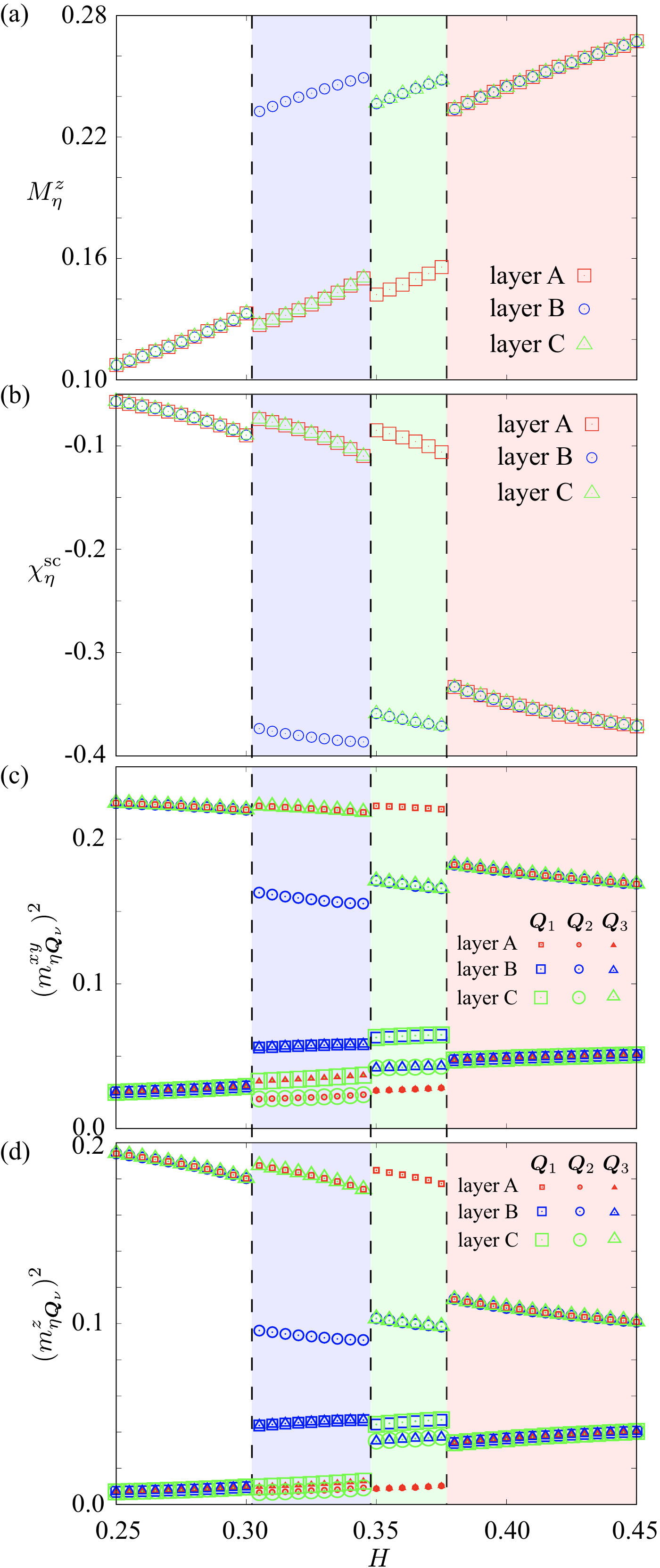} 
\caption{
\label{fig: mq_pSkX}
$H$ dependences of (a) $M^z_\eta$, (b) $\chi^{\rm sc}_\eta$, (c) $(m^{xy}_{\eta \bm{Q}_\nu})^2$, and (d) $(m^{z}_{\eta \bm{Q}_\nu})^2$ for $\kappa=0.76$. 
The vertical dashed lines represent the phase boundaries between the different phases.
}
\end{center}
\end{figure}

\begin{figure*}[t!]
\begin{center}
\includegraphics[width=0.85 \hsize ]{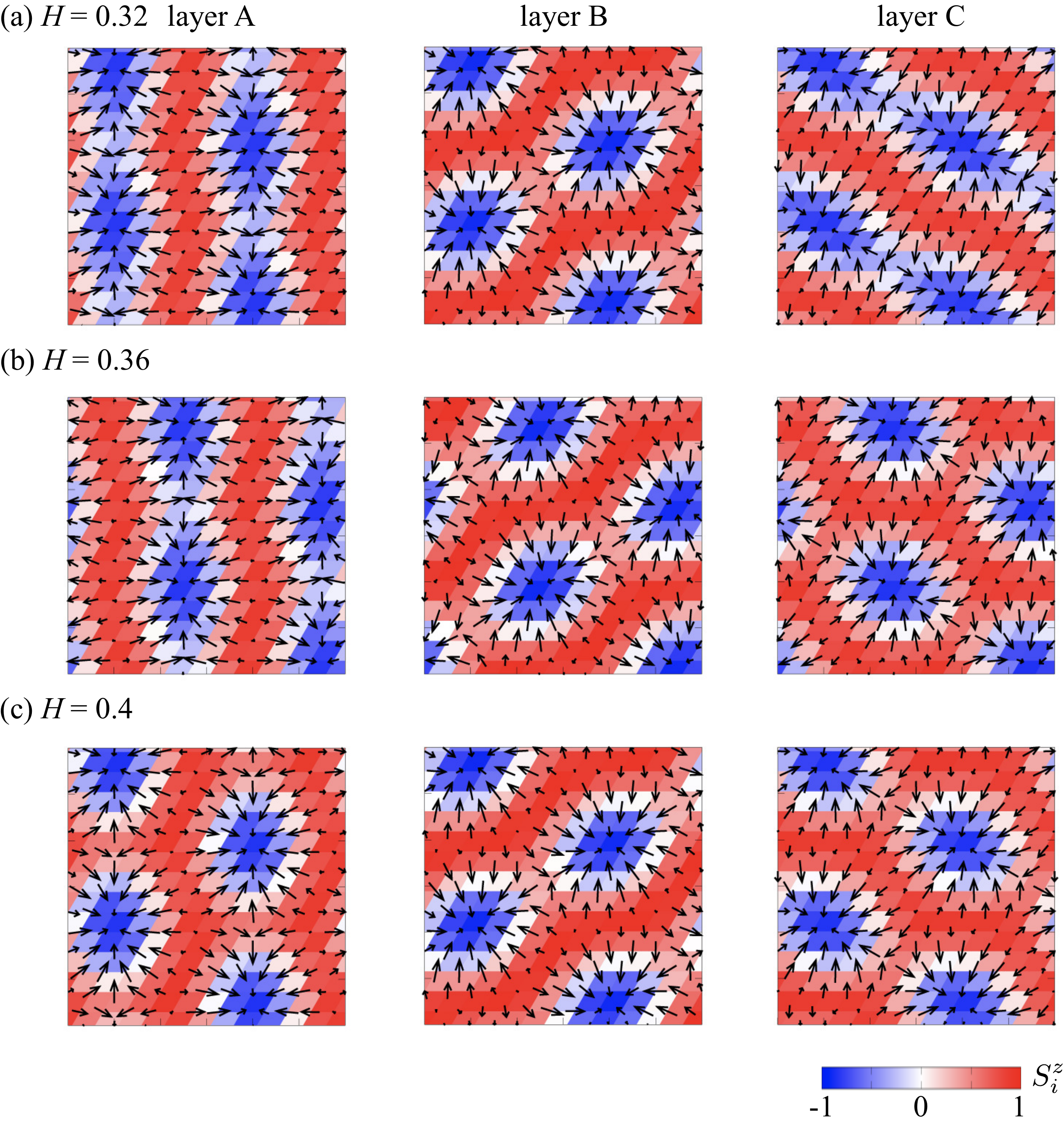} 
\caption{
\label{fig: spin_pSkX}
Real-space spin configurations of (a) the f-SkX-I at $H=0.32$, (b) the f-SkX-II at $H=0.36$, and (c) the SkX at $H=0.4$ on the layers A (left), B (middle), and C (right) for $\kappa=0.76$. 
The color represents the $z$ component of the spin moment, and the arrows stand for the $xy$ components. 
}
\end{center}
\end{figure*}

Next, we consider the stability of the SkX while changing $\kappa$ and $H$ for fixed $J_{\parallel}$. 
We here choose $J_{\parallel}=0.2$ so that the SkX is not stabilized at $\kappa=0$, as shown by the phase diagram in the $\kappa$-$H$ plane in Fig.~\ref{fig: PD_ratiodep}, where the color plot represents the scalar chirality $\chi^{\rm sc}$. 
For $\kappa=0$, only the triple-$Q$-I state appears before changing into the fully-polarized state against $H$ in the phase diagram, where the spiral plane changes from the $xz$ or $yz$ plane to the $xy$ plane similar to the result in Fig.~\ref{fig: mq_J=0.1} in Sec.~\ref{sec: Weak interlayer coupling}. 

While increasing $\kappa$, the SkX is stabilized in the intermediate magnetic field, as shown in Fig.~\ref{fig: PD_ratiodep}; the SkX appears for $\kappa \gtrsim 0.31$ and its region extends while increasing $\kappa$ (It is noted that the system with $\kappa=1$ possesses the threefold rotational symmetry in addition to the threefold screw symmetry). 
In the low-field region, the 1$Q$ state is stabilized instead of the 3$Q$-I state and it remains stable in the high-field region. 
This 1$Q$ state shows a similar spin configuration to that in the 3$Q$-II state; the difference lies in the presence of the small amplitudes of $(\bm{m}_{{\rm A}\bm{Q}_2})^2$, $(\bm{m}_{{\rm A}\bm{Q}_3})^2$, $(\bm{m}_{{\rm B}\bm{Q}_1})^2$, $(\bm{m}_{{\rm B}\bm{Q}_3})^2$, $(\bm{m}_{{\rm C}\bm{Q}_1})^2$, and $(\bm{m}_{{\rm C}\bm{Q}_2})^2$ in the 3$Q$-II state. 
Thus, the instability tendency when increasing $\kappa$ is similar to that when increasing $J_{\parallel}$, which indicates that the SkX tends to be more stabilized for relatively large $\kappa$ and $J_{\parallel}$. 

Remarkably, we find that two different types of the SkXs in the phase diagram, which are denoted as fractional SkX I (f-SkX-I) and fractional SkX II (f-SkX-II) in Fig.~\ref{fig: PD_ratiodep}. 
We show the spin- and chirality-related quantities in the vicinity of the region where these two phases are stabilized in Fig.~\ref{fig: mq_pSkX}. 
The data for $0.25 \leq H \lesssim 0.3$ correspond to the 3$Q$-I state, while that for $0.38 \lesssim H \leq 0.45$ correspond to the SkX, both of which are characterized by the dominant peaks at $m^{xy}_{{\rm A}\bm{Q}_1}$, $m^{xy}_{{\rm B}\bm{Q}_2}$, $m^{xy}_{{\rm C}\bm{Q}_3}$, $m^{z}_{{\rm A}\bm{Q}_1}$, $m^{z}_{{\rm B}\bm{Q}_2}$, and $m^{z}_{{\rm C}\bm{Q}_3}$ and the subdominant peaks at the other $m^{xy}_{\eta\bm{Q}_{\nu}}$ and $m^{z}_{\eta\bm{Q}_{\nu}}$, as shown in Figs.~\ref{fig: mq_pSkX}(c) and \ref{fig: mq_pSkX}(d). 
The difference between these two phases is found in the skyrmion number: the 3$Q$-I state has no skyrmion number, while the SkX has the skyrmion number of $-1$, as discussed in Sec.~\ref{sec: Instability toward skyrmion crystal phase}. 

The f-SkX-I and f-SkX-II phases exhibit a coexistence feature of the 3$Q$-I state and the SkX. 
As shown in Figs.~\ref{fig: mq_pSkX}(a) and \ref{fig: mq_pSkX}(b), the behaviors of $M_\eta^z$ and $\chi_\eta^{\rm sc}$ for layer B (layers B and C) resemble the Sk, while those for layers A and C (layer A) resemble the 3$Q$-I state in the f-SkX-I (f-SkX-II) phase. 
This indicates that the SkX spin texture is realized for one layer (two layers) in the f-SkX-I (f-SkX-II) state. 
Indeed, the layer-dependent skyrmion numbers are obtained in these two states; the f-SkX-I and f-SkX-II exhibit a set of the skyrmion number as $(n^{\rm (A)}_{\rm sk}, n^{\rm (B)}_{\rm sk}, n^{\rm (C)}_{\rm sk})=(0,-1,0)$ and $(0,-1,-1)$, respectively, where $n^{\rm (\eta)}_{\rm sk}$ represents the skyrmion number in layer $\eta$. 
It is noted that any permutation between $(n^{\rm (A)}_{\rm sk}, n^{\rm (B)}_{\rm sk}, n^{\rm (C)}_{\rm sk})$ is allowed owing to the symmetry. 
Thus, the f-SkX-I and f-SkX-II have the fractional skyrmion numbers of $-1/3$ and $-2/3$ per magnetic unit cell, respectively. 
This is why we call these states the fractional SkXs. 
The layer-dependent spin textures in these two states are shown in the real-space spin configuration. 
We present the spin configurations in each layer in the f-SkX-I at $H=0.32$ in Fig.~\ref{fig: spin_pSkX}(a), the f-SkX-II at $H=0.36$ in Fig.~\ref{fig: spin_pSkX}(b), and the SkX at $H=0.4$ in Fig.~\ref{fig: spin_pSkX}(c). 
One finds that both the f-SkX-I and f-SkX-II consist of different layers, whose spin textures are similar to those in the SkX and the 3$Q$-I state.
As the skyrmion numbers are different among the 3$Q$-I state, f-SkX-I, f-SkX-II, and SkX, the results provide a realization of functional magnetic materials, whose topological properties are switched by an external magnetic field, in the nonsymmorphic systems with the screw axis.

\section{Summary}
\label{sec: Summary}

We have investigated the instability toward the SkX in the nonsymmorphic system without the threefold rotational symmetry but with the threefold screw symmetry. 
By performing the simulated annealing for the layered spin model, where each layer is connected by the threefold screw symmetry, we have shown that the interplay between the interlayer exchange coupling and the anisotropic momentum-resolved interaction plays an important role in stabilizing the SkX from zero to finite magnetic fields. 
We also found two layer-dependent SkXs with the fractional skyrmion numbers in a magnetic unit cell under an external magnetic field. 
Our result indicates a possibility to realize the SkX in the nonsymmorphic system without a solely rotational symmetry within the same plane, which leads to extending the scope of the SkX-hosting materials. 
The targeting lattice structures are represented by the space groups $P3_1$ (\# 144), $P3_2$ (\# 145), $P3_1 12$ (\# 151), $P3_1 21$ (\# 152), $P3_2 12$ (\# 153), and , $P3_2 21$ (\# 154).
Moreover, the space groups having sixfold and fourfold screw symmetry, such as $P4_1$ (\# 76), $P4_122$ (\# 91), $P6_1$ (\# 169), and $P6_122$ (\# 178), are also candidates hosting the SkXs. 
As the magnetic materials belonging to these space groups are rare as listed in MAGNDATA, magnetic structure database~\cite{gallego2016magndata}, e.g., $X$Fe$_3$(BO$_3$)$_4$ ($X=$ Dy, Tb, Ho, and Y)~\cite{ritter2007magnetic,ritter2008magnetic,ritter2012magnetic} and BaCu$_3$V$_2$O$_8$(OD)$_2$~\cite{Boldrin_PhysRevLett.121.107203} under $P3_1 21$ (\# 152), it is highly desired to investigate similar materials belonging to the above space groups to observe the SkX in the nonsymmorphic system with the screw axis.

In addition, the present results provide a potential realization of the multiple-$Q$ spin crystals other than the SkX, such as the meron-antimeron~\cite{Lin_PhysRevB.91.224407,yu2018transformation,hayami2018multiple,kurumaji2019skyrmion,Hayami_PhysRevB.104.094425,chen2022triple}, hedgehog~\cite{kanazawa2017noncentrosymmetric,fujishiro2019topological,Kanazawa_PhysRevLett.125.137202,grytsiuk2020topological,Ishiwata_PhysRevB.101.134406,Okumura_PhysRevB.101.144416,Mendive-Tapia_PhysRevB.103.024410,Shimizu_PhysRevB.103.054427,Kato_PhysRevB.104.224405}, vortex~\cite{Momoi_PhysRevLett.79.2081,Kamiya_PhysRevX.4.011023,Wang_PhysRevLett.115.107201,Marmorini2014,Hayami_PhysRevB.94.174420,Solenov_PhysRevLett.108.096403,Ozawa_doi:10.7566/JPSJ.85.103703,takagi2018multiple,yambe2020double}, and bubble crystals~\cite{lin1973bubble,Garel_PhysRevB.26.325,takao1983study,Hayami_PhysRevB.93.184413,Su_PhysRevResearch.2.013160,seo2021spin}, in the nonsymmorphic systems with the screw axis. 
As these spin textures are stabilized by the anisotropic exchange interaction and the multiple-spin interaction that have not been taken into account in the present model, further exotic spin textures that have never been reported in the single-layer system might be expected in the nonsymmorphic system with the layer degrees of freedom. 
Such a discovery of new magnetic phases will be useful to explore intriguing electronic structures~\cite{heinze2011spontaneous,Bergmann_PhysRevB.86.134422,von2015influence,hanneken2015electrical,Ozawa_PhysRevLett.118.147205,Kubetzka_PhysRevB.95.104433,Kathyat_PhysRevB.103.035111,Christensen_PhysRevX.8.041022,Hayami_PhysRevB.104.144404,Hayami_PhysRevB.105.024413} and quantum transport phenomena~\cite{Nagaosa_RevModPhys.82.1539,tokura2018nonreciprocal} characteristics of the multiple-$Q$ states in nonsymmorphic systems, which will be left for future study. 

\begin{acknowledgments}
This research was supported by JSPS KAKENHI Grants Numbers JP19K03752, JP19H01834, JP21H01037, and by JST PRESTO (JPMJPR20L8).
Parts of the numerical calculations were performed in the supercomputing systems in ISSP, the University of Tokyo.
\end{acknowledgments}

\bibliographystyle{apsrev}
\bibliography{ref}

\begin{thebibliography}{152}
\bibitem[{com({\natexlab{b}})}]{comment_screw}
\bibinfo{note}{The threefold screw symmetry is slightly broken in the
  intermediate field region, as found in Fig.~\ref{fig: mq_J=0.1}(b).}
\end{thebibliography}

\end{document}